\documentclass[11pt,a4paper]{article}
\usepackage{longtable}
\usepackage{amsmath}
\usepackage{amssymb}
\usepackage{graphicx}
\usepackage{bm}

\newcommand*{\bea}{\begin{eqnarray}}
\newcommand*{\eea}{\end{eqnarray}}
\newcommand*{\be}{\begin{equation}}
\newcommand*{\ee}{\end{equation}}

\newcommand{\bma}{\begin{pmatrix}}
\newcommand{\ema}{\end{pmatrix}}

\usepackage[square,comma,sort&compress,numbers]{natbib}
\title{Dynamical Mass Generation in Wick Cutkosky Model}
\author{Tajdar Mufti \footnote{tajdar.mufti@gmail.com, tajdar.mufti@lums.edu.pk} \\ Lahore University of Management Sciences\\ Opposite Sector U, D.H.A, Lahore Cantt., 54792, Pakistan}
\begin{document}
\maketitle
\begin{abstract}
Studying extent of dynamical generation of mass in a quantum field theory, which may have non-perturbative attributes, is an essential step towards complete understanding of the theory. The phenomenon has historic relevance to interactions including fermions. However, as searches for further fundamental scalars continue, inquiring into the possibility of dynamical generation of mass becomes crucial for scalars with masses much lower than the electroweak scale. This paper addresses Yukawa interaction between a complex doublet field, named the Higgs for convenience, and a scalar singlet field studied in terms of correlation functions and the dynamical masses produced in the parameter space of the model. An unorthodox approach is adopted to extract the correlation functions and dynamical masses in the model. The study is conducted using the method of Dyson Schwinger equations for the Yukawa coupling $10^{-6} \leq \lambda \leq 2.0$ in GeVs and cutoff $10 \leq \Lambda \leq 100$ in TeVs. The scalar masses are found to be around afew MeVs while the Higgs masses are found to be smaller than 1 MeV. The Higgs propagators are found much more sensitive on the coupling than the scalar propagators. The vertices are relatively more stable against the cutoff used in comparison to the propagators and dynamical masses. The model is found to be stable for the cutoff above $100$ TeV. The critical coupling exists in the model and is expected to be in the vicinity of $\lambda=10^{-6}$ GeV. No indications of triviality is found in the model.
\end{abstract}
\section{Introduction}
Importance of the standard model (SM) \cite{Schwartz:2013pla,Barnett:1996yz} as the theory of low energy phenomenology for particle interactions can not be overstated after the discovery of Higgs boson \cite{pdg,Maas:2017wzi,Carena:2002es} at the LHC \cite{Aad:2012tfa,Chatrchyan:2012xdj,MalbertionbehalfoftheCMSCollaboration:2018eqs}. The discovery also opens a possibility of existence of other fundamental scalars (or pseudo-scalars) in nature, particularly the ones to describe the beyond SM physics, such as Supersymmetry \cite{Kane:2018oax,Haber:1989xc,Gunion:2002zf,Martin:1997ns,Aaboud:2017leg}, cosmic inflation \cite{Bezrukov:2013fka,Linde:2005ht,Mukhanov:2005sc,Enckell:2018kkc,Guth:1980zm,Ferreira:2017ynu,Hakim:1984oya,Linde:1993cn}, and dark matter \cite{Athron:2017kgt,Lee:2017qve,Bento:2000ah,Bertolami:2016ywc,Munoz:2017ezd} physics. As there exists a wide range of masses in nature from a few orders less than $1$ eV to close to $200$ GeVs, the possibilities of masses for further scalars is a an open question.
\par
One of the main reasons to expect the Higgs was to render the electroweak gauge bosons and several other fundamental particles massive. However, it is a well known fact that the lightest quarks do not direly require Higgs mechanism as they can also generate masses via QCD interactions \footnote{There also exist various propositions in literature related to this matter, see for example \cite{Hansson:2014cva}.}. With no information regarding the masses of further scalars, studying extent of dynamical generation of masses (DMG) \cite{Brauner:2005hw,Maris:1999nt,Fischer:2003rp,Aguilar:2005sb,Bowman:2005vx,Aguilar:2010cn,Cloet:2013jya,Mitter:2014wpa,Binosi:2016wcx,Ayala:2006sv,Libanov:2005vu,Benes:2008ir} via scalar interactions in various models naturally becomes an important aspect of the scalar sector.
\par
An element of ambiguity in studies related to new physics is that it is not truly known if the new physics is accessible through the approach of perturbation. Furthermore, DMG is a non-perturbative phenomenon by definition. Hence, non-perturbative methods, such as that of Dyson Schwinger Equations (DSEs) \cite{Schwinger:1951ex,Schwinger:1951hq,Swanson:2010pw,Roberts:1994dr,Rivers:1987hi}, become necessary tool to investigate the underlying physics.
\par
However, as is the case with almost every non-perturbative approach, the method of DSEs have a limitation that there may be more unknown correlation functions than the number of considered DSEs. To cope with, it is not uncommon to use ansatz and truncations \cite{Schwinger:1951ex,Schwinger:1951hq,Swanson:2010pw,Roberts:1994dr,Ayala:2006sv} which may effect the correlation functions. Thus, beside studying the physics, simpler (than the SM, for instance) models offer an extremely important playground for implementing and studying unconventional numerical approaches. A model containing the Yukawa interaction vertex is a highly interesting avenue to study scalar interactions due to its relevance to particle physics phenomenology.
\par
This paper is a continuation of studies of a variant of Wick Cutkosky model \cite{Mufti:2018xqq,Darewych:1998mb,Sauli:2002qa,Efimov:2003hs,Nugaev:2016uqd,Darewych:2009wk}, and addresses the phenomenon of DMG due to a Yukawa interaction between the Higgs field, which preserves SU(2) symmetry, and a (real) singlet scalar field using the approach of DSEs \cite{Schwinger:1951ex,Schwinger:1951hq,Swanson:2010pw,Roberts:1994dr,Rivers:1987hi}. There are several benefits such an study offers. Firstly, the model serves as a useful avenue to study how two mutually interacting fields with different symmetries and under the same renormalization condition dynamically acquire masses for various cutoff values. Secondly, understanding features, such as existence of critical coupling and phase structure, in a model containing Yukawa interaction may provide valuable information in searches for new physics. Furthermore, as mentioned above, a relatively simpler interaction offers a test ground to employ uncustomary approaches which could later be used in richer theories.
\par
The model is studied for (bare) coupling values $10^{-6} \leq \lambda \leq 2.0$ in GeVs \footnote{Bare couplings are taken in GeV in order to keep the study in the perspective of electroweak physics. Furthermore, no peculiarities of the Wick Cutkosky model is used.} and cutoff values at $10$ TeV, $50$ TeV, and $100$ TeV. By the definition of DMG, both fields are massless unless they interact.
\par
A significant amount of studies involving scalar interactions is with inclusion of four point interactions in the realm of renormalizable quantum field theories. It includes both the four point self interactions of the Higgs or scalar fields, or the vertex formed by two scalar and two Higgs fields. However, these three types of interactions can also be formed by the three point Yukawa interaction mentioned above. Thus, in this paper only the simpler Yukawa interaction is considered. Inclusion of further vertices are to be reported somewhere else \cite{tajdar:2018lat1} using a different non-perturbative approach \cite{Ruthe:2008rut,Gattringer:2010zz}.
\par
The $\phi^{4}$ theory \cite{Hasenfratz:1988kr,Gliozzi:1997ve,Weber:2000dp} is found to be trivial \cite{Jora:2015yga,Aizenman:1981zz,Weisz:2010xx,Siefert:2014ela,Hogervorst:2011zw}. However, despite being a (complex doublet) scalar field, Higgs interactions with gauge bosons in the Yang-Mills-Higgs theory \cite{Maas:2013aia,Maas:2014pba} is not found to render the theory trivial. Hence, it is implicitly assumed here that the model considered is not trivial.
\par
This paper encompasses study of the theory in terms of correlation functions and dynamically generated masses using two DSEs for the two field propagators without the truncations \footnote{Here, using limited number of equations has not been taken as a truncation as it will generally be the case for an interacting theory.} or ansatz mentioned above. The DSEs for the two field propagators are used to extract the correlation functions in a numerically controlled environment. The only hard constraints are renormalization conditions, which are typical of renormalizable quantum field theories, and a restriction on the vertex to restrict violent local fluctuations. The equation for least square error is taken as the third equation for the system of three unknown correlation functions. The details are mentioned in the next section.
\section{Technical Details}
The Euclidean version of the Lagrangian \footnote{Bare masses are included in the Lagrangian for the sake of clarity.}, along with the counter terms \cite{Das:2008zze}, is given by \footnote{A considerable part of technical details can also be found in other reports. The details are kept in their entirety for the sake of self-sufficiency.}
\begin{equation} \label{Lagrangian}
 \begin{split}
L= & (1+A)\delta^{\mu \nu} \partial_{\mu} h^{\dagger} \partial_{\nu} h + (m_{h}^{2} + B ) h^{\dagger} h + (1+ \alpha) \frac{1}{2} \delta^{\mu \nu} \partial_{\mu} \phi \partial_{\nu} \phi + \\  
& \frac{1}{2} (m_{s}^{2} + \beta) \phi^{2}+ \lambda (1+\frac{C}{\lambda})  \phi h^{\dagger} h
\end{split}
\end{equation}
with h as the Higgs fields with SU(2) symmetry and $\phi$ a real scalar singlet field. $A$, $B$, $C$, $\alpha$, and $\beta$ are coefficients of the counter terms, and $\lambda$ is the three point interaction coupling. The DSEs for scalar and the Higgs propagators are, respectively, given by
\begin{equation} \label{spr1:dse}
\begin{split}
 S(p) ^{-1}  = & (1+\alpha)p^{2} + m_{s}^{2}+ \beta+ \\
 & \lambda (1+\frac{C}{\lambda})  \int \frac{d^{4}q}{(2\pi)^{4}} H^{ik}(q) \Gamma^{kl}(q,p-q,-p) H^{li}(q-p)
 \end{split}
\end{equation}
\begin{equation} \label{hpr1:dse}
\begin{split}
 H^{ij}(p)^{-1} = & \delta^{ij} ((1+A)\ p^{2} + m_{h}^{2} + B )\  + \\ 
 & 2 \lambda(1+\frac{C}{\lambda}) \int \frac{d^{4}q}{(2\pi)^{4}} S(q) \Gamma^{ik}(-p,p-q,q) H^{kj}(q-p)
 \end{split}
\end{equation}
where $\Gamma^{kl}(u,v,w)$ is the three point Yukawa interaction vertex of the Higgs, the Higgs bar, and the scalar singlet fields with momenta $u$, $v$, and $w$, respectively. The Higgs and the Higgs bar \footnote{Throughout the paper, Higgs bar is referred to $h^{\dagger}$.} fields have indices $k$ and $\l$, respectively. $S(p)$ and $H^{ij}(p)$ are scalar propagator and the Higgs propagator, respectively. Setting $m_{s}=0$ and $m_{h}=0$, and introducing the following definitions of $B$ and $\beta$, respectively,
\begin{equation} \label{B:def}
B = 2 \lambda (1+A) (1+\alpha) \sigma_{h}
\end{equation}
\begin{equation} \label{beta:def}
\beta = 2 \lambda (1+A) (1+\alpha) \sigma_{s}
\end{equation}
the dynamical squared masses of the Higgs and scalar fields, respectively, assume the following definitions \footnote{Since $\sigma_{h}$ and $\sigma_{s}$ can take any suitable value during computation, the definitions do not impose any constraints on the dynamical squared masses.}.
\begin{equation} \label{sqmh:def}
m^{2}_{h,d} = 2 \lambda (1+A) (1+\alpha) \sigma_{h}
\end{equation}
\begin{equation} \label{sqms:def}
m^{2}_{s,d} = 2 \lambda (1+A) (1+\alpha) \sigma_{s}
\end{equation}
The vertex is defined as
\begin{equation} \label{vtx:def}
\Gamma^{ik}_{r} = (1+\frac{C}{\lambda}) \Gamma^{ik} = (1+A) (1+\alpha) \tilde{\Gamma}^{ik}
\end{equation}
Hence, we have the following equations \footnote{The definitions introduced in equations \ref{sqmh:def}-\ref{vtx:def} results in a multiplicative constant in each of the equations \ref{spr2:dse} and \ref{hpr2:dse} which facilitate in implementation of the renormalization conditions.},
\begin{equation} \label{spr2:dse}
\begin{split}
S(p) ^{-1}  = & (1+\alpha) (\ p^{2} + 2 \lambda (1+A) \sigma_{s} + \\
 & \lambda (1+A) \int \frac{d^{4}q}{(2\pi)^{4}} H^{ik}(q) \tilde{\Gamma^{kl}}(q,p-q,-p) H^{li}(q-p) )\ 
\end{split}
\end{equation}
\begin{equation} \label{hpr2:dse}
\begin{split}
H^{ij}(p)^{-1} = & (1+A) (\ \delta^{ij} (\ p^{2} + 2 \lambda (1+\alpha) \sigma_{h})\  + \\
 & 2 \lambda (1+\alpha) \int \frac{d^{4}q}{(2\pi)^{4}} S(q) \tilde{\Gamma^{ik}}(-p,p-q,q) H^{kj}(q-p) )\
\end{split}
\end{equation}
The renormalization conditions \cite{Roberts:1994dr} for the propagators are given below \footnote{The renormalization point is chosen at 1 GeV.}.
\begin{equation} \label{hpr:ren_condition}
H^{ij}(p)  |_{p^{2}=1}= \frac{\delta ^{ij}}{p^{2}} |_{p^{2}=1}
\end{equation}
\begin{equation} \label{spr:ren_condition}
S(p)  |_{p^{2}=1}= \frac{1}{p^{2}} |_{p^{2}=1}
\end{equation}
As mentioned below, the quantities $1+A$ and $1+\alpha$ are calculated directly from the renormalization conditions. If the propagators are dominated by tree level contribution, the quantities $(1+A)$, $(1+\alpha)$, and $(1+\frac{C}{\lambda})$ approach 1. As a result, the vertex becomes tree level dominated, see equation \ref{vtx:def}. Hence, the definition of the vertex in \ref{vtx:def} is justified. It is also the reason that in this study computations for all the parameters start with tree level expression for the Higgs propagators, and $\tilde{\Gamma}^{ik}=\lambda$ in magnitude. Since $\tilde{\Gamma}^{ik}$ changes during a computation, the vertex remains without any strict constraints.
\par
No renormalization condition was explicitly imposed on the vertex \footnote{It was found that introducing any renormalization condition on the vertex produces abnormal discontinuities in the vertices for very low coupling. This behavior was taken as a sign of over-constrained system.}. The symmetries in the DSEs are numerically implemented while at the same time only the flavor diagonal Higgs propagators are assumed to be non-vanishing in order to maintain the similarity between the Higgs propagator and its tree level structure.
\par
Interpolation is performed on scalar propagator to implement the renormalization condition \ref{spr:ren_condition}. Since the two DSEs, \ref{spr2:dse} and \ref{hpr2:dse} are nonlinearly coupled, $m^{2}_{h,d} \geq 0$ and $m^{2}_{s,d} \geq 0$ are imposed in order to suppress the artifacts due to interpolations.
\par
As there is no DSE or ansatz used for the Yukawa vertex, three steps are taken to ensure that the resulting correlation functions, particularly the vertex, are stable. Firstly, a condition is locally imposed on the vertex that it never exceeds an order of magnitude relative to its neighboring momentum points in the 4-momentum (Euclidean) spacetime. The constraint is similar to the well known Lipschitz condition abundantly used in literature \cite{Gupal1980,Gu2001,Aronsson1967,Zevin2011,Ezquerro2016}. The constraint is introduced to keep the vertex from fluctuating violently. Secondly, instead of calculating local deviations, sum of squared errors is calculated, which leads to implementation of least squares method with the equation given below.
\begin{equation} \label{lseeq}
\begin{split}
\int dp (\ H^{ij}(p)^{-1} - & (\ (1+A) (\ \delta^{ij} (\ p^{2} + 2 \lambda (1+\alpha) \sigma_{h})\  +  2 \lambda (1+\alpha) \\
 & \int \frac{d^{4}q}{(2\pi)^{4}} S(q) \tilde{\Gamma^{ik}}(-p,p-q,q) H^{kj}(q-p) )\ )\ )\ ^{2} =0
\end{split}
\end{equation}
It certainly slows the computations but it is found enormously helpful in improving stability of the correlation functions. It also serves as the third equation and, hence, fulfilling the requirement of the third equation for the three correlation functions to be calculated \footnote{There are also two renormalization conditions imposed, which fix $1+A$ and $1+\alpha$. Hence, the system of equation can be solved to find unique solutions.}. Furthermore, the solution becomes an optimization problem in which $\ref{lseeq}$ is to be minimized.
\par
Lastly, the Higgs propagators are expanded in the polynomial form given below.
\begin{equation} \label{hpr:coeff1}
 H^{ij}(p)= \delta^{ij} \frac{1}{c(p^{2}+d+f(p))}
\end{equation}
with $f(p)$ defined as
\begin{equation} \label{hpr:func1}
f(p)=\frac{\displaystyle \sum_{i=0}^{N} a_{i} p^{2i}}{\displaystyle \sum_{l=0}^{N}b_{l}p^{2l}}
\end{equation}
where $a_{i}$, $b_{l}$, $c$, and $d$ are parameters to be updated during a computation. Such a parameterization brings certain advantages. Firstly, the procedure of renormalization is significantly faster than the alternative approaches which may involve interpolations. Secondly, there is a certain correspondence between the self energy contribution and the expansion in equations \ref{hpr2:dse} and \ref{hpr:func1}, which renders the vertex a certain form. Hence, a stable vertex is achieved whose fluctuations mostly depend upon the resolution among momentum values.
\par
The computation proceeds as follows: It starts with tree level structures for the Higgs propagators and the vertex \footnote{$c=1$, and $a_{i}$, $b_{l}$, and $d$ are set to zero in equations \ref{hpr:coeff1} and \ref{hpr:func1}. The vertex is set at the coupling value.}, and the parameters $\sigma_{s}$ and $\sigma_{h}$ are set to zero. The scalar propagator takes the values from its DSE, see equation \ref{spr2:dse}. First, $\sigma_{s}$ is calculated using Newton Raphson's method. The criterion for acceptance of update is minimization in equation \ref{lseeq}, i.e. the value of the integral in the equation decreases. If an update does not decrease the value, it is not accepted. The same method is used to update the Higgs propagator, the vertex, and $\sigma_{h}$. Update of $\sigma_{s}$ is followed by that of $\sigma_{h}$ parameter.
\par
Next is the update of the parameters in equation \ref{hpr:coeff1} for the Higgs propagators. Lastly, the function $\tilde{\Gamma^{ik}}$ is updated at each momentum value while preserving the symmetry imposed by the DSEs. During updating of each of the above mentioned quantities, scalar propagator is calculated from equation \ref{spr2:dse}. The quantities $1+A$ and $1+\alpha$ are calculated during each update and calculation of the Higgs and scalar propagators, respectively.
\par
Hence, as a computation proceeds, $\sigma_{s}$ and $\sigma_{h}$ deviate from their starting value, while the Higgs propagators and the vertex numerically deviate from their tree level structures \footnote{During updates, if a parameter or the vertex at a momentum point does not improve the result by decreasing the above mentioned error, the quantity is not updated and algorithms move on to the next point of momentum (for the case of the vertex) or paramter.}. The computation ends when either there is no further improvement in the sum of squared errors or it has reached a value below the preset value for the sum of squared errors. The minimum error is preset at $10^{-20}$.
\par
It was found that changing the sequence of updates of the parameters or the correlation functions does not effect the results within machine precision. It was taken as the definition of uniqueness through out the study.
\par
The model is studied with interactions taking place on a plane immersed in the 4 dimensional Euclidean spacetime. Gauss quadrature algorithm is used for numerical integration.
\par
The algorithms are developed in C++ environment and ROOT, CERN is used for constructing the presented diagrams.
\section{Correlation Functions}
\subsection{Field Propagators}
\begin{figure}
\centering
\includegraphics[width=\linewidth]{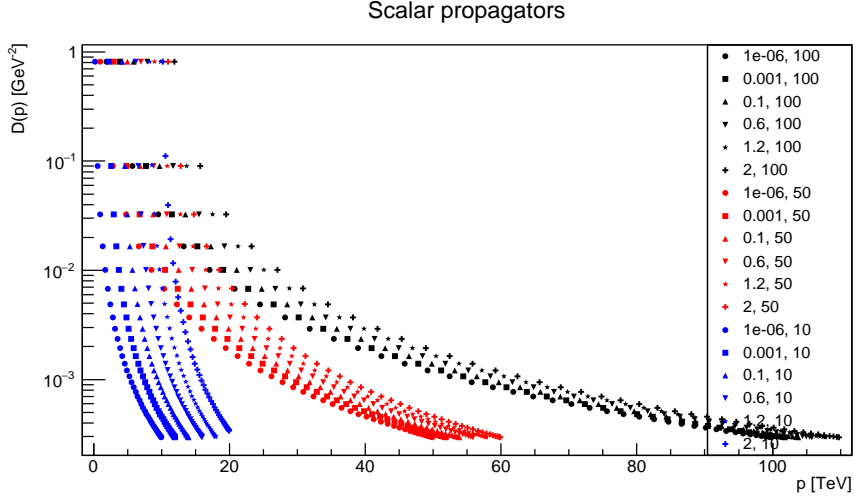}
\caption{\label{fig:sprs1} Scalar propagators for different couplings $\lambda$ (in GeVs) and cutoff values $\Lambda$ (in TeVs), shown as ($\lambda$,$\Lambda$) in the legend, are plotted. Starting with the highest value of coupling constant for a fixed cutoff, each subsequent scalar propagator with lower coupling constant is displaced by 2 TeV momentum in the figure.}
\end{figure}
\begin{figure}
\centering
\includegraphics[width=\linewidth]{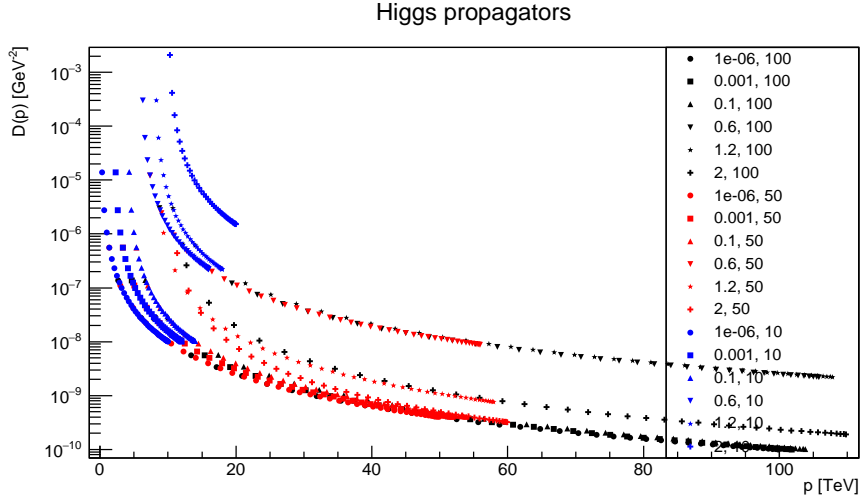}
\caption{\label{fig:hprs1} Higgs propagators for different couplings $\lambda$ (in GeVs) and cutoff values $\Lambda$ (in TeVs), shown as ($\lambda$,$\Lambda$) in the legend, are plotted. Starting with the highest value of coupling constant for a fixed cutoff, each subsequent Higgs propagator with lower coupling constant is displaced by 2 TeV momentum in the figure.}
\end{figure}
The scalar and the Higgs propagators are shown in figures \ref{fig:sprs1} and \ref{fig:hprs1}, respectively. For the case of scalar propagators, the most interesting feature is their strong (qualitative as well as quantitative) similarity for various couplings. It points towards the possibility that the model favors a certain narrow range of dynamical scalar masses which, in an extreme case, can even be a particular value characteristic to the model. In fact, it was the Higgs propagator which was expected to be stable due to the reason that Higgs mass was found to be less sensitive \cite{Gies:2017zwf}. However, the presence of cutoff effects in the propagators is severe enough to expect cutoff dependence of scalar mass. The effects become milder for higher cutoff which is an indication of absence of coupling dependence as the cutoff value reaches the order of hundreds of TeV, see figure \ref{fig:sprs1}. Overall, the propagators are found enhanced as the coupling increases for all cutoff values. It is an indication of the role of the quantity $1+\alpha$ in equation \ref{spr2:dse}.
\par
In contrast to the scalar propagators, the Higgs propagators posses different features. First of all, they are sensitive to coupling as well as the cutoff values. For the coupling values considerably lower than 1.0 GeV, Higgs propagators are found to be suppressed. The propagators are enhanced as the coupling rises to the vicinity of 1.0 GeV. However, signs of similar suppression are observed once again for further higher coupling values. This change in behavior is peculiar since it takes place in the vicinity of $\lambda=0.6$ GeV which is close to the fourth root of quartic self interaction coupling for Higgs. Furthermore, as is the case for scalar propagators, there are cutoff effects for higher coupling values.
\par
Both propagators loose dependence on cutoff in TeVs and coupling lower than $10^{-3}$ GeV coupling values. It is the first clue that the masses might have approached a value characteristic to the model.
\subsection{Higgs-scalar vertices}
\begin{figure}
\centering
\includegraphics[width=\linewidth]{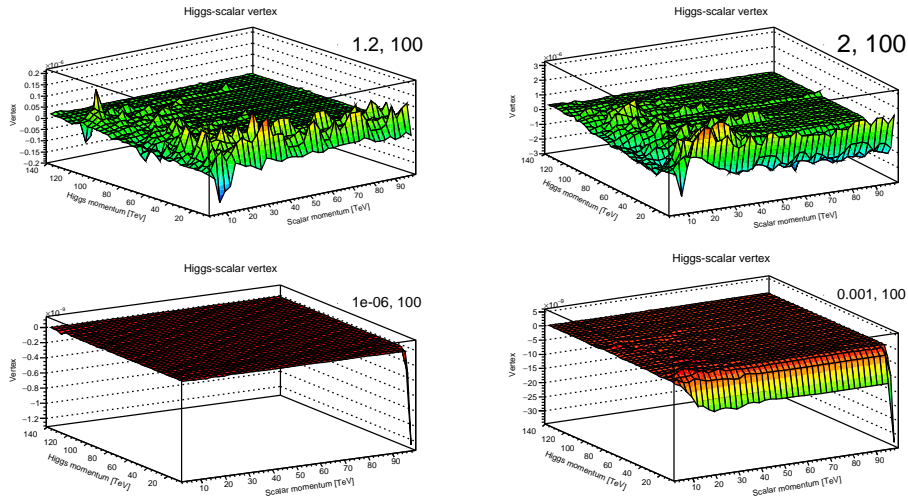}
\caption{\label{fig:v100} Yukawa interaction vertices between Higgs and scalar fields are shown for various couplings (in GeV) with cutoff at $100$ TeV.}
\end{figure}
\begin{figure}
\centering
\includegraphics[width=\linewidth]{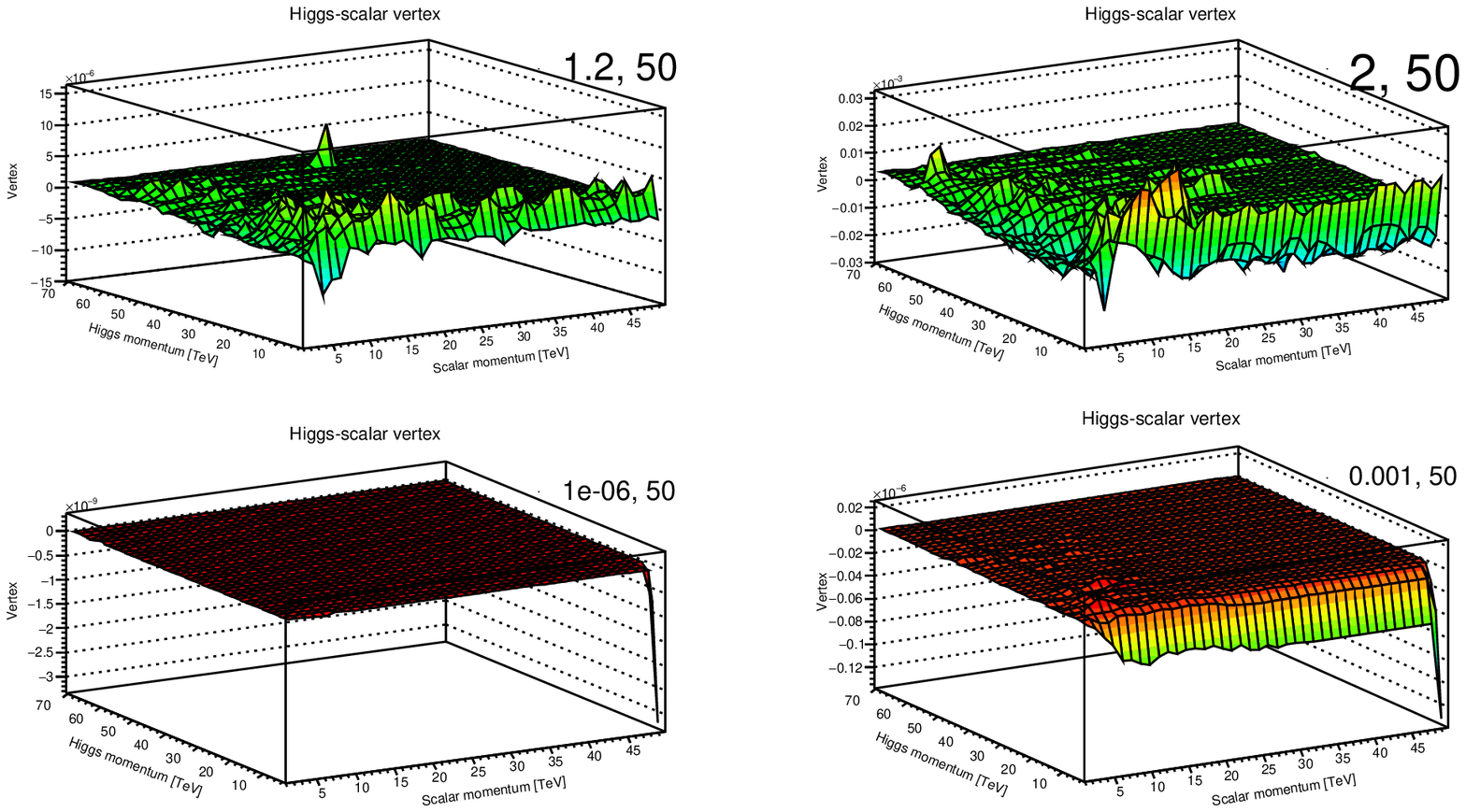}
\caption{\label{fig:v50} Yukawa interaction vertices between Higgs and scalar fields are shown for various couplings (in GeV) with cutoff at $50$ TeV.}
\end{figure}
\begin{figure}
\centering
\includegraphics[width=\linewidth]{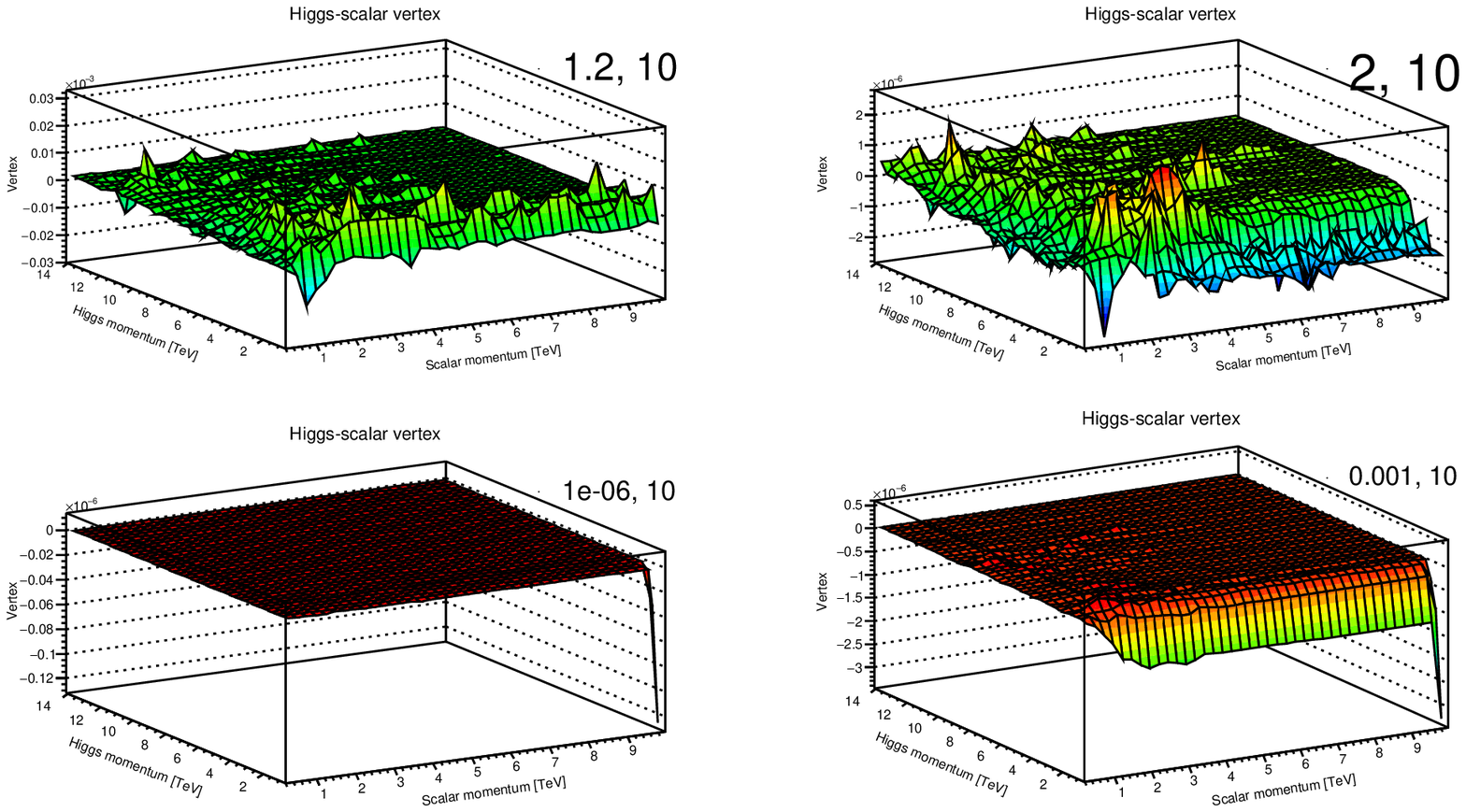}
\caption{\label{fig:v10} Yukawa interaction vertices between Higgs and scalar fields are shown for various couplings (in GeV) with cutoff at $10$ TeV.}
\end{figure}
The Yukawa interaction vertices, defined in equation \ref{vtx:def}, are shown in figures \ref{fig:v100}, \ref{fig:v50}, and \ref{fig:v10} for various coupling values with cutoff at $100$ TeV, $50$ TeV, and $10$ TeV, respectively \footnote{The presence of fluctuations is due to lesser resolution among the field momenta.}.
\par
Firstly, the vertex is found to have qualitative dependence on the field momenta, see figures \ref{fig:v100} to \ref{fig:v10}. A peculiar observation is the similarity of the vertices for $\lambda \leq 10^{-3}$ GeV irrespective of the cutoff. It suggests that the physics (at around or) below this coupling value does not change much.
\par
For higher couplings the vertices do not resume their behavior at low coupling values, which was observed for the Higgs propagators. It is a sign that for higher values of coupling, the dynamical masses may behave differently over coupling. Since such a qualitative dependence is not in harmony with scalar propagators, as they have weak dependence on the coupling,  it suggests that the Higgs' role may be relatively more prominent than that of scalar field in the phenomenon of dynamical mass generation.
\par
The vertex is also found to have weak cutoff dependence at higher couplings. It supports the speculation that it may be the masses whose cutoff effects translate to the propagators, see equations \ref{spr2:dse} and \ref{hpr2:dse}. 
\par
Qualitative dependence on field momenta is a clear indication that the theory is not a trivial theory and the deviations in the Higgs propagators are more than a mere multiplicative constant to the corresponding tree level structure due to the contribution by self energy term, see equation \ref{hpr2:dse}. It indicates that the extracted correlation functions may differ from the studies with assumtion that the vertex is fixed at a certain value value for all field momenta.
\section{Dynamical Renormalized Masses}
\begin{figure}
\centering
\includegraphics[width=\linewidth]{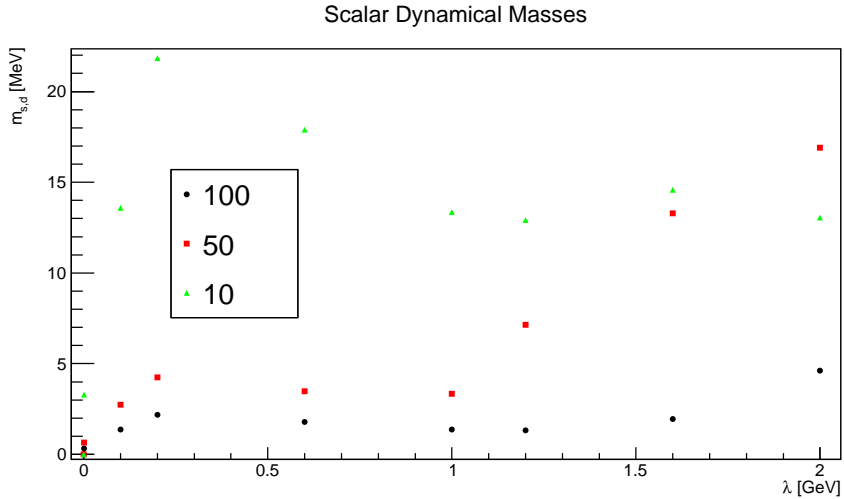}
\caption{\label{fig:sdynmass} Dynamically generated masses of scalar field $m_{s}$ as a function of Yukawa coupling $\lambda$ at different cutoff values (in TeVs) are shown.}
\end{figure}
\begin{figure}
\centering
\includegraphics[width=\linewidth]{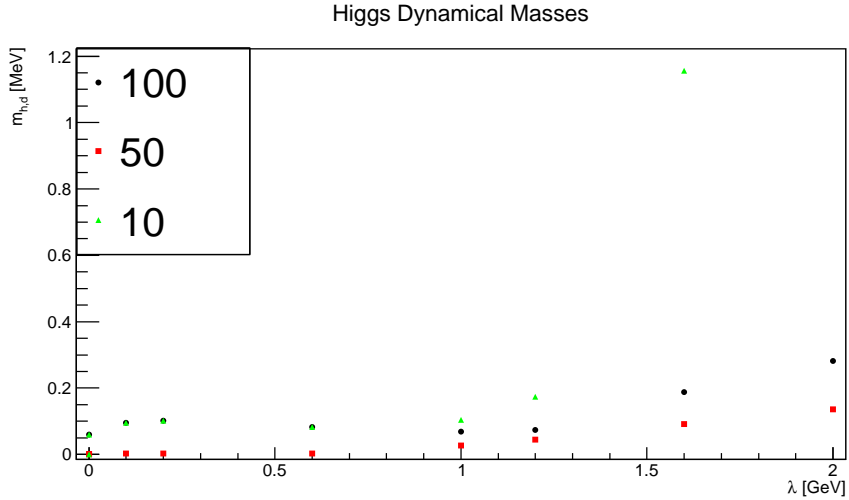}
\caption{\label{fig:hdynmass} Dynamically generated squared masses of the Higgs field $m_{s}$ as a function of Yukawa coupling $\lambda$ at different cutoff values (in TeVs) are shown.}
\end{figure}
The dynamical masses are plotted in figures \ref{fig:sdynmass} and \ref{fig:hdynmass} for scalar and the Higgs fields, respectively. An immediate observation is strong cutoff effects on masses of both fields. For the case of DMG, the renormalized masses do not have any contribution from tree level value. It is also the reason that the cutoff effects appear vividly in the figures \ref{fig:sdynmass} and \ref{fig:hdynmass} and can not hide behind the bare mass. Indication of stability in the masses ensues only above 100 TeVs.
\par
For the case of scalar mass, it remains within afew MeVs for most of the coupling values at higher cutoff. This observation is not entirely unexpected as the stability of the scalar propagators has already suggested it. However, peculiar to the model is the magnitude of the mass which is in the vicinity of the lightest quarks \cite{pdg}.
\par
The Higgs dynamical mass is found to be less than 0.5 MeV, significantly lower than the scalar mass. It demonstrates how dynamics of the fields with different symmetries differ from each other in a model despite the fact that statistically they are of the same type of fields.
\par
Dynamical masses for both fields remains non-vanishing for $2.0 \leq \lambda \leq 10^{-3}$ but are found to be zero at $\lambda = 10^{-6}$ GeV. Hence, based upon their dependence on the coupling, it is concluded that the critical coupling may exist in the vicinity of $\lambda = 10^{-6}$ GeV.
\par
Despite the difference in magnitudes, the cutoff at which the masses tend to stabilize is over a hundred TeV. In comparison to the QCD physics, this observation can be attributed to the simplicity of the model which does not contain any but a Yukawa interaction.
\section{Conclusion}
The current study is an addition to exploration of Wick Cutkosky model using the method of DSEs with a different approach. It is found that even a Yukawa interaction can dynamically generate scalar masses of the magnitude close to the already existing fundamental particles.
\par
The propagators as well as the masses can be differentiated for the two fields. Hence, the study also serves as a demonstrator of how two fields of the same nature but different symmetries in the theory menifest themselves. The vertices are least effected by the cutoff. 
\par
As is the case with QCD interactions, the model is found to be producing masses in MeVs which serves as another limitation on the extent of dynamical mass generation. However, an important difference is that the masses stabilize over most of the coupling values for cutoff above 100 TeVs. Hence, the model serves as a reminder of the remarkable strength and diversity of QCD interactions which could be competed, though in a restricted sense, with a Yukawa interaction only at relatively much higher cutoff values.
\par
It was found that there is indeed a critical coupling for the model at around $10^{-6}$ GeV. 
\par
model is an addition to the observation that the Higgs interaction with other fields does not render the theory trivial. Thus, it begs for further investigation of the scalar sector in the models richer than the one studied here.
\section{Acknowledgments}
I am deeply indebted to Dr. Shabbar Raza and Dr. Rizwan Khalid for a number of valuable discussions during this endeavor. I would also like to express my deepest gratitude to Professor Johan Hansson for his critical advices and inspiration by some of his work.
\par
This work was supported by Lahore University of Management Sciences, Pakistan.
\bibliographystyle{plain}
\bibliography{bib}
\end{document}